# Use of Hamilton's canonical equations to rectify Newton's corpuscular theory of light: A missed opportunity

by

Robert J. Buenker[*]
Fachbereich C-Mathematik und Naturwissenschaften
Bergische Universität Wuppertal
Gaussstrasse 20
D-42119 Wuppertal, Germany

**Abstract**

The erroneous prediction of the speed of light in dispersive media has been looked upon historically as unequivocal proof that Newton's corpuscular theory is incorrect. Examination of his arguments shows that they were only directly applicable to the momentum of photons, however, leaving open the possibility that the cause of his mistake was the unavailability of a suitable mechanical theory to enable a correct light speed prediction, rather than his use of a particle model. It is shown that Hamilton's canonical equations of motion remove Newton's error quantitatively, and also lead to the most basic formulas of quantum mechanics without reference to any of the pioneering experiments of the late nineteenth century. An alternative formulation of the wave-particle duality principle is then suggested which allows the phenomena of interference and diffraction to be understood in terms of statistical distributions of large populations of photons or other particles.

---

## I. Introduction

On the basis of his corpuscular theory of optics, Newton predicted that light travels faster in a normally dispersive medium than in free space.[1] When Foucault measured the speed of light in water nearly 150 years after the publication of *Opticks*, it was clear that the opposite is the case, and as a result, belief in the particle model was virtually abandoned in favor of the wave theory of optical phenomena. The trend had been in this direction anyway since the interference experiments of Young[2] were reported in 1802, which gave strong support to the superposition principle introduced by Huygens in the late seventeenth century, and any remaining question about the wave theory of light was seemingly eliminated in 1864 with the publication of Maxwell's electromagnetic theory. Yet only 40 years later, Einstein's interpretation of the photoelectric effect[3] in terms of light quanta with $E = \hbar\omega$ disturbed this consensus. Then in 1923 Compton was able to construct a quantitative theory for the scattering of x-rays by valence electrons of atoms[4] by using conservation of energy/momentum arguments which were quite consistent with Newton's corpuscular theory.[1] In recognition of these developments, Lewis[5] coined the word "photon" to denote a light quantum, and in succeeding years single-photon detectors have become commonplace in the modern physics laboratory. In spite of this, Newton's erroneous prediction of the speed of light in dispersive media is thought to be irrefutable evidence against a strictly particle theory of light.[6] Instead, the concept of duality[7] is widely used to describe the fact that all matter seems to behave as if it is composed of either waves or particles, depending on the type of experiment to which it is subjected.

## II. Light refraction and quantum mechanics

In view of the significance that has been attached over the years to Newton's light speed prediction, it is interesting to examine his line of reasoning in arriving at this position. His essential argument was that the fact that the ratio of the sines of the angles of incidence $\Theta_1$ and refraction $\Theta_2$ (Fig. 1) always have the same ratio for a given pair of transparent media (Snell's Law) is consistent with a force acting at the interface between the media and in the direction normal to it. Light travels in a straight line within any homogeneous medium, indicating that there are no unbalanced forces except those at interfaces between different media. According to his Second Law, this means that the component of the photon momentum parallel to any such interface is a constant of motion, as expressed by the following set of proportionalities:

$$p_1/p_2 = n_1/n_2 = \sin\Theta_2/\sin\Theta_1, \qquad (1)$$

where $p_i$ is the magnitude of the photon's momentum in a given medium and $n_i$ is the corresponding refractive index. In other words, p is proportional[8] to n, which means, for example, that the photon momentum for yellow light in water is 1.33 times greater than in free space. Consistent with this position, he argued that the angle of refraction of light of a given wavelength near the surface of the earth under otherwise equivalent conditions only depends on its angle of incidence as it enters the upper atmosphere, that is, it is independent of how the air pressure varies along the way. Newton did not use the term "momentum" explicitly in presenting his arguments, but preferred to use "velocity" instead.[9] There can be no doubt that in so doing, he was simply assuming that the inertial mass m = p/v of the corpuscles is the same in all media. This raises the question, however, of whether it was not exactly this supposition that led him astray, rather than the particle model itself.

Mechanical theory was not sufficiently developed in Newton's lifetime to provide an accurate description of the dynamics of particles in the presence of external fields. The situation was greatly improved 130 years after publication of *Opticks*, however, when Hamilton introduced his canonical equations of motion. It then became clear that the velocity v could be determined with knowledge of the momentum dependence of the total energy E, specifically as

$$v = dE/dp \qquad (2)$$

in the simplest case (in fact, this equation can easily be deduced from the Second Law and the definition of energy). Newton had shown that white light is decomposed into its component colors when it passes into a dispersive medium, and he clearly associated this phenomenon with the varying accelerations experienced by particles of light of different color. Since white light travels great distances from the sun and the stars without undergoing an analogous divergence, it follows by the same reasoning that the speed of light c has the same constant value for all photons in free space. Roemer had been able to give a respectably accurate value for c based on his observations of the moons of Jupiter in 1635. There is apparently no record of Hamilton having done so, but if one simply sets v=c in eq. (2) and integrates, the result is Einstein's well-known special relativity equation[10] for photons in free space,

$$E = pc, \qquad (3)$$

whereby the constant of integration is set to zero and serves as part of the definition of E. It would also have been possible to see from eq. (2) that by assuming m to be invariant, the standard formula of $E = p^2/2m$ leads to a different result ($v = p/m$) which is inconsistent with the constancy of c. This finding might simply have caused confusion in 1834, but today we know that it is indicative of the failure of nonrelativistic theory to describe the motion of photons.

In order to compute the velocity of light in dispersive media it is necessary to obtain a suitable generalization of the $E = pc$ relation in free space. Newton's theory had shown that p is proportional to n (see Fig. 1), so the first step is simply to replace p by p/n. Then one needs a comparable relationship to express the dependence of energy on the nature of the refracting medium. Newton was well aware that the color of light rays does not change as they pass from one medium to another, only their direction, so the simplest assumption would have been and still remains that E is completely independent of n. In other words, there is no exchange of energy between the photons and a given transparent medium through which they travel. On this basis the desired generalization of eq. (3) becomes:

$$E = pc/n . \qquad (4)$$

Applying Hamilton's equations of motion, i.e. eq. (2), yields

$$v = c/n - pc/n^2 \, dn/dp = c/n - kc/n^2 \, dn/dk . \qquad (5)$$

If one ignores the more complicated derivative term for the time being, it is seen that on this basis the speed of light in water does not increase with n as Newton concluded, but is actually inversely proportional to it, exactly as the wave theory of light had predicted. By assuming that the momentum of the photons is proportional to n, one is led by Hamilton's equations to conclude that their velocity will change in the opposite direction. Had Newton or anyone else taken this approach prior to 1850, the shock value of Foucault's experiment would have been completely eliminated and it would have been recognized that measurements of the speed of light in dispersive media are *incapable* of providing a definitive answer as to whether the particle or the wave theory of light is incorrect.

Moreover, the best measurements of the speed of light in dispersive media that have been carried out over the years[11-13] indicate that eq. (5) is exact. To see this, however, it is necessary to know the relation between the photon momentum p and wavelength $\lambda$ and wave vector k = $2\pi/\lambda$. Newton had reported careful measurements[14] of "Intervals of Fits of easy Reflexion and Transmission" in the rings he observed at glass-air interfaces, and noted in his Prop. XVII that they were in the same proportion as the sines of incidence and refraction in different media. One hundred years later, Young[15] noted that these intervals were simply one-half of the wavelength $\lambda$ and was able to compute accurate values for various colors of light based on Newton's measurements. Because of eq. (1), this means that both p and n are inversely proportional to lambda and therefore proportional to k, so that

$$p = \hbar k, \qquad (6)$$

where $\hbar$ is simply a constant of undetermined magnitude with units of angular momentum. This relation was apparently first used by Compton in 1923[4,16] but when substituted in eq. (5), it leads to the expression on the far right as derived from an otherwise strictly corpuscular model of light.

One can only speculate why Newton did not deduce eq. (6) from his Prop. XVII, but the probable reason is that he considered it a meaningless relation since it combines quantities, namely momentum and wavelength, from, in his view, mutually contradictory theories. It is possible to carry the point further along by examining the fundamental relation for the phase velocity in the wave theory of light,

$$v_p = \omega/k = c/n . \qquad (7)$$

The phase velocity is never measured in dispersive media, as will be discussed subsequently, so this relation simply expresses the fact that the frequency $\omega$ is independent of n, whereas k is proportional to it. Comparison with eq. (4) shows that $v_p$ is numerically equal to E/p, from which one can conclude from eq. (6) that

$$E = \hbar\omega. \qquad (8)$$

This is the famous relation introduced by Planck[17] in 1900 in his theory of blackbody radiation which was later used by Einstein in his interpretation of the photoelectric effect.[3] It follows directly from Newton's corpuscular theory when Hamilton's eq. (2) is used to compute the speed of light in dispersive media and thus arguably could have been known as early as 1834, well before the dawn of the quantum age. Had this occurred, the aim of future experiments would have been much more clearly defined than it was historically, namely simply to measure the value of $\hbar$ in eqs. (6,8) to as high an accuracy as possible.

**III. Assumptions of the wave theory**

Having shown that the exact dependence of the speed of light on refractive index n can be derived from the particle model, it is instructive to consider how the same result is obtained in the wave theory. Application is made of Rayleigh's theory of sound[18] and its explanation of how beats arise when two waves of equal amplitude but slightly differing frequency and wavelength interfere. Using trigonometric relationships it was shown that two distinct wave motions result, that of the carrier (wavelets) propagating with the phase velocity obtained from the average of the two frequencies and wavelengths, and that of an envelope or wave group characterized by their differences, $\Delta\omega$ and $\Delta k$. The group velocity $v_g$ is thus the ratio of the latter two quantities or $d\omega/dk$ in the limit of infinitesmal differences. This derivative can be evaluated from eq. (7) in the case of light refraction[19] and the result is identical with the observed light speed relation given in eq. (5).

To justify this approach, it is necessary to assume that whenever monochromatic light falls upon a dispersive medium, waves of slightly differing $\omega$ and k are always formed and it is the speed of the resulting wave groups which is determined in experiments such as Foucault's. It should be noted, however, that the corresponding $\Delta\omega$ and $\Delta k$ quantities have never been observed experimentally. This is explained by claiming that these differences are simply too small to be measured, but this means that both the period and the wavelength of the wave groups are essentially infinitely long. At the same time, the frequency and wavelength of the monochromatic light are observed, but their corresponding (phase) velocity is also never measured in refractive media. This situation is unlike any of the classical applications of Rayleigh's theory to sound and water waves. When two musical instruments are slightly out of tune, both the average tone and the characteristic beat frequency are easily audible. When a rock is dropped into a pond, both wavelets and wave groups are clearly visible. Arguing that $\Delta\omega$ and $\Delta k$ are too small to be observed for light waves still raises the question as to why the associated group velocity should be measured if one has to wait an *infinitely long time* to observe a complete wave group unit. In short, the supposed perturbation of monochromatic light waves in refractive media may be purely hypothetical.

The $d\omega/dk$ of the wave theory is exactly equal to $dE/dp$ by virtue of eqs. (6,8), and so the exact light velocity expression of eq. (5) is obtained from it. In the particle theory of light, there is no need to argue that photons do not all have the same velocity in a given medium to arrive at the same result.[20,21] Formally, one simply needs to know the dependence of their energy on momentum to evaluate their speed by means of eq. (2). The ratio of the speeds of a single photon in free space and in a refractive medium, which is traditionally referred to as the group index of refraction $n_g$, can be obtained directly from this expression as

$$n_g = c/v = d(pc)/dE = d(nE)/dE \qquad (9)$$
$$= n + E\, dn/dE = n + \omega\, dn/d\omega,$$

when used in conjunction with eqs. (4,8). In recent experiments[22,23] quantum interference effects have been exploited to produce very steep variations in the refractive index of gaseous media with frequency in order to reduce the speed of light to as low as 17 m s$^{-1}$. Despite the quite large values of $dn/d\omega$ attained in these investigations, the refractive index itself remains very close to unity throughout, and thus according to eq. (1) the corresponding photon momenta are virtually unchanged relative to their values in free space. The effect of the electromagnetically induced transparency in these experiments is therefore to increase the inertial mass of a single photon by seven orders of magnitude without greatly altering either its energy or momentum.

## IV. Statistical interpretation of duality

Realisation that a proper treatment of the motion of photons does account quantitatively for the measured variations of the speed of light in refractive media removes one of the most fundamental objections against the atomistic theory of matter proposed by Democritus and his followers in ancient Greece and later espoused by Newton in his corpuscular model. The quantum mechanical concept of wave-particle duality[7], by contrast, holds that matter behaves as particles in some experiments but as waves in others, and as such can be viewed as a compromise between two theories which were traditionally thought to be mutually exclusive. Nonetheless, it is argued that because of the Heisenberg uncertainty principle, there can be no such thing as a perfectly localized particle. In this view, a "particle" is simply a particularly localized wave packet (a quantized state of the electromagnetic field), so from a purely philosophical point of view, duality is heavily slanted toward the wave theory of matter.

It is possible to make a different interpretation of quantum mechanical duality, however, one which is far more consistent with atomistic principles:

**Some experiments are so precise (photoelectric and Compton effects, the refraction of light and single-photon counting) that they reveal the elementary nature of matter in terms of particles, while others (interference and diffraction as the primary examples) are only capable of giving information about the statistical distribution of particles in space and time.**

The latter distribution is given in the Born interpretation[24] as $\Psi\Psi^*$, the absolute square of the quantum mechanical wave function. Accordingly, the uncertainty principle merely states that if all that is known about a collection of indistinguishable particles is that they each possess the same momentum, then quantum mechanics can only say that they are *no more likely* to be found in one location than in any other at any given time. A snapshot taken of a large ensemble of such entities will always appear the same, even though one knows that from one moment to the next there is a constant exchange of particles in any given location, since they are all moving with the same known velocity.

In an interference or diffraction experiment, if the intensity of the beam is small and detection is made with a device such as a photographic plate, the distribution observed early in the counting procedure will vary significantly from one trial to another. If the experiment is continued for a sufficiently long period of time in each case, however, the pattern of detected objects will stabilize to agree completely with quantum mechanical $\Psi\Psi^*$ predictions. Moreover, if the intensity is lowered far enough, single particles can always be detected one at a time,[25] which is probably the strongest experimental argument for a purely atomistic theory of matter. In this view, a single atom, molecule, photon, or electron is *not* vibrating with a definite frequency and wavelength. Rather, the k and the $\omega$ in eqs. (6) and (8) are the parameters in $\Psi$ that specify the statistical distribution that many identical particles of this kind possess as an ensemble. One needs a significantly large number of such objects in order to obtain sufficiently reliable values for $\omega$ and k from experimental observations, whereby the period of time over which these measurements are made is not a key factor in such determinations. As in other applications of statistics, the resulting distributions may be quite inadequate for predicting the behavior of individuals, but they provide an unerring guide for trends within very large populations. Accordingly, the wave packet bears the same relationship to the particle as the histogram does to a member of a sample whose statistical distribution it represents. The latter is a real object, whereas the former is only a mathematical abstraction. A light wave is certainly real, but in analogy to an ocean wave containing many water molecules, it is a collective body whose elementary constituents are single photons.

It should also be mentioned that there is a simple explanation for the polarization of light in the particle model. Wigner[26] has shown that because of relativistic considerations, although the angular momentum quantum number of photons is J=1, only their $M_J$=+1 and -1 components are ever observed, and therefore that the two polarizations of light can be distinguished on this basis. Furthermore, vacuum fluctuations in quantum electrodynamics can also be understood in a qualitative manner by assuming that the condensed Bose-Einstein state of light consists of real photons of zero energy[27] that are unobservable because of eqs. (6,8).

Two other common objections to such a statistical interpretation of quantum mechanics need to be mentioned briefly. In a number of key applications, it is found that there is a finite probability for particles to exist in regions where they are classically forbidden, such as for the harmonic oscillator or in tunneling processes. When a measurement is carried out in the classically forbidden region, the value expected for the potential energy V from classical mechanics will be obtained, but in the process the wave function is changed along with its total energy[28], so that nothing prevents the corresponding kinetic energy T from being positive and therefore classically allowed. By the same token, prior to the measurement, there is no justification for computing either T or V by classical means because the wave function is only an eigenstate of the total Hamiltonian (energy), so one also cannot be certain that T must be negative outside the classically allowed region in this case.

Finally, it has also been argued that the results of the Young double-slit experiment are inconsistent with such a statistical interpretation[29,30] of quantum mechanics. It is assumed in essence that a single photon does not have the capacity to go through two open slits on the way to the detector, but that a wave does. A thorough analysis of the observed data advises that greater caution be exercised on this point, however. If one also employs very small intensities in this experiment and detection is again made with a photographic plate or its equivalent placed behind the screen which contains the slits, it is found that the resulting distribution accumulates point by point in a thoroughly random fashion. Stopping the counting at a relatively early stage always produces a series of distinct points on the photographic plate, and not a continuous wave distribution. If only photons (or electrons or atoms or molecules) are counted in the statistics for the case when both slits of the Young apparatus are open, eventually the well-known interference pattern will result. This, in turn, is quite different from what is measured when only those events are counted which occur when just one of the holes is open. More importantly, adding the results for the two opposing individual single-slit experiments gives a distribution which is quite distinct from that which results when both slits are open simultaneously. Moreover, the holes can be opened and closed many times during the passage of the light or electron beam to the detector, but all that matters is the conformation of the slits at the time of actual passage through the screen. In other words, at the time the matter is emitted from the source, the intensity pattern which will eventually be observed after sufficient accumulation at the detector is not predictable with certainty unless one knows what the conformation of slits will be at the instant of subsequent passage through the screen. In short, there seems to be no means of understanding this series of observations satisfactorily without invoking some undetected object or interaction which ultimately determines the outcome of the experiment. Under the circumstances, it seems fairer to admit that the nonlocal character[31] of the interaction between system and measuring device in this case is not fully explained by either the wave or particle models. It is more properly considered as a separate issue, and, just as the light dispersion experiments discussed first, should not be used to rule out either of them as a theory for the elementary composition of matter.

## V. Conclusion

Newton saw a clear application of his Second Law in the light refraction experiments he had carried out. The irony is that by failing to distinguish between trends in momentum and velocity, he was led to a prediction that eventually damaged the scientific reputation of his *Opticks* and directed attention to the competing theory of matter he so disliked. In the present work, it has been shown that the correct expression for the velocity of light in refractive media is obtained by applying Hamilton's canonical equations of motion to Newton's corpuscular theory. Moreover, the key quantum mechanical equations, $E = \hbar\omega$ and $p = \hbar k$, can be deduced from Newton's own observations on this basis, without reference to any of the pioneering experiments carried out at the end of the 19th century. What needs to be emphasized from this exercise is not just speculation about how much earlier key theoretical principles might have been deduced, but more importantly, the fact that it shows that standard arguments which have hitherto been brought against the atomistic theory of matter because of Newton's optical experiments are totally without foundation. Recognition of this point should lead to a thorough examination of both the positions that perfectly localized particles are not ruled out by the uncertainty principle, and that experiments such as interference and diffraction can indeed be explained on the basis of Newton's corpuscular model.

**Acknowledgements**

This work was supported in part by the Deutsche Forschungsgemeinschaft within the Schwerpunktprogramm "Theorie relativistischer Effekte in der Chemie und Physik schwerer Elemente."

**Figure Caption**

**Fig. 1**. Schematic diagram showing the refraction of light at an interface between air and water. The relation between the angles of incidence $\Theta_1$ and refraction $\Theta_2$ in terms of the refractive indices $n_i$ (Snell's Law) of the two media was viewed by Newton as a clear application of his Second Law of kinematics, according to which the component of the photon momentum $p_i$ parallel to the interface must be conserved.

**Fig.1**

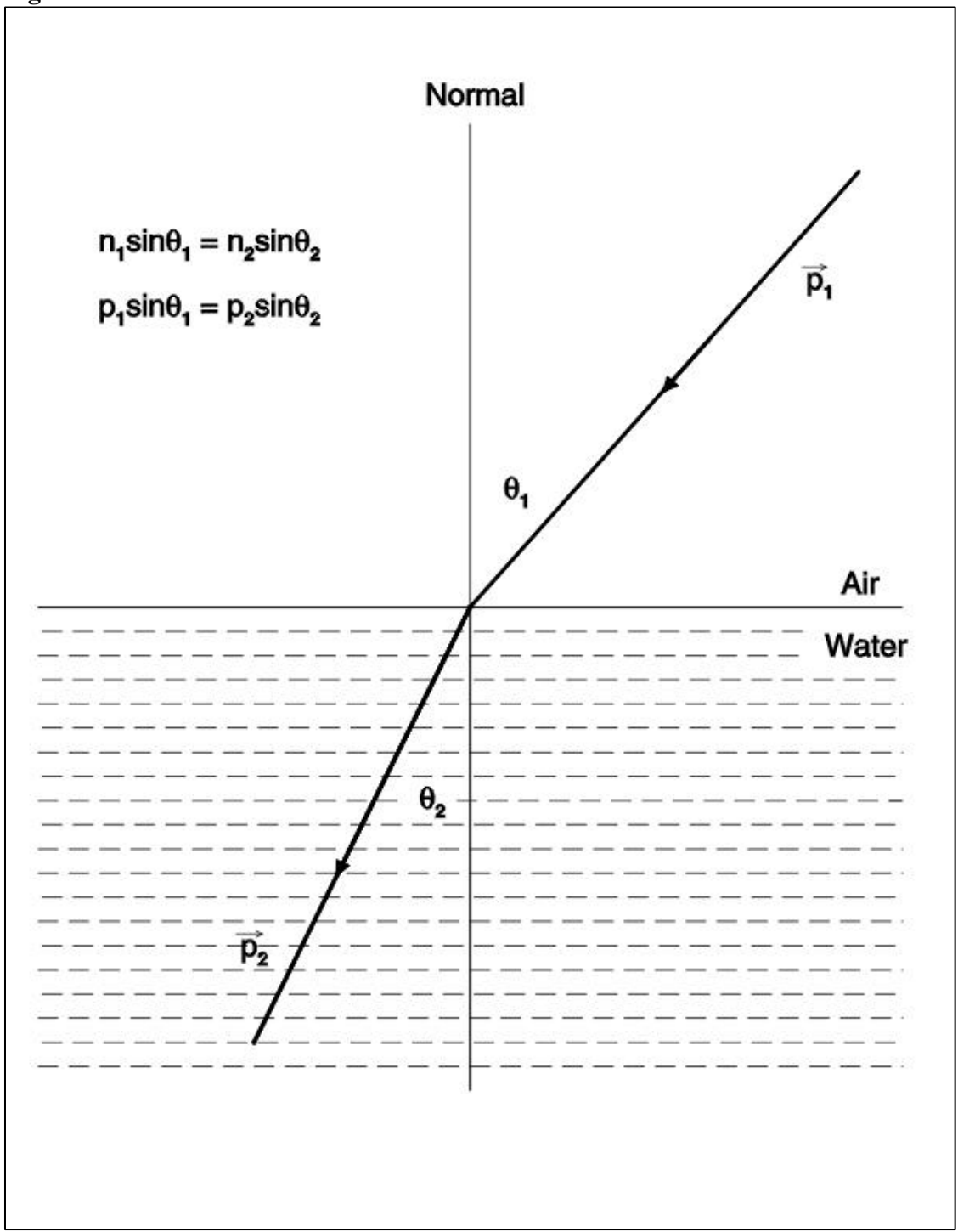